\documentclass[prapplied,amssymb,amsmath,superscriptaddress,twocolumn]{revtex4-2}

\usepackage{graphicx,dcolumn,mathptmx,gensymb,color}
\usepackage[T1]{fontenc}

\newcommand{\hl}[1]{\textcolor[RGB]{255,0,0}{#1}}
\renewcommand{\hl}[1]{#1} 

\graphicspath{{figures/}}

\begin{document}

\title[]{A continuous, \hl{sub-Doppler-cooled} atomic beam interferometer for inertial sensing}

\author{J.~M. Kwolek}
\affiliation{Optical Sciences Division, U.S. Naval Research Laboratory, Washington, D.C. 20375}
\author{A.~T. Black}
\email{adam.black@nrl.navy.mil}
\affiliation{Optical Sciences Division, U.S. Naval Research Laboratory, Washington, D.C. 20375}

\date{\today}

\begin{abstract}
We present the first demonstration of an inertially sensitive atomic interferometer based on a continuous, rather than pulsed, atomic beam at sub-Doppler temperatures in three dimensions. We demonstrate 30\% fringe contrast in continuous, inertially sensitive interference fringes at interrogation time $T=6.7~\mathrm{ms}$ and a short-term phase measurement noise of $530~\mathrm{\mu rad /\sqrt{Hz}}$ as inferred from interference measurements. Atoms are delivered to the interferometer by a cold-rubidium source that produces a high flux of atoms at temperature $\leq15~\mathrm{\mu} K$ in three dimensions while reducing near-resonance fluorescence in the downstream path of the atoms. We describe the optimization of the interrogating Raman beams to achieve high contrast, and validate interferometer operation through comparison with measurements by commercial accelerometers. We further provide a demonstration of zero-dead-time phase-shear readout of atom interferometer phase, achieving a measurement rate up to 160~Hz. This demonstration lays the groundwork for future gyroscope/accelerometer sensors that measure continuously, with both high bandwidth and high sensitivity, and on dynamic platforms.
\end{abstract}

\maketitle

\section{Introduction}

In light-pulse atom interferometers \cite{kasevich_atomic_1991,kasevich_measurement_1992,geiger_high-accuracy_2020, narducci_advances_2021}, coherent interaction with optical fields causes atoms to propagate in a superposition of spacetime trajectories that interfere at the output of the interferometer. Applications of light-pulse atom interferometry to measurement of acceleration \cite{mcguinness_high_2012,lautier_hybridizing_2014,cheiney_navigation-compatible_2018}, rotation rate \cite{gustavson_rotation_2000,durfee_long-term_2006,savoie_interleaved_2018,avinadav_rotation_2020}, gravity \cite{peters_high-precision_2001, bidel_compact_2013, wu_gravity_2019}, and gravity gradients \cite{snadden_measurement_1998,biedermann_testing_2015} have been demonstrated on numerous occasions, and future applications such as gravitational wave detection \hl{\cite{tino_atom_2011,dimopoulos_gravitational_2009,canuel_exploring_2018,abe_matter-wave_2021}} are under development.

Interferometry in continuous atomic beams \cite{riehle_interferometry_1992,gustavson_precision_1997, durfee_long-term_2006, xue_continuous_2015} has long been studied as a method of achieving high bandwidth, high signal-to-noise ratio, and elimination of aliasing, each of which presents challenges in pulsed cold-atom interferometers. These features are advantageous for sensing of dynamically varying quantities in applications such as inertial navigation or gravimetry on moving platforms, or in vibrationally noisy environments. These continuous-beam, or ``spatial-domain,'' interferometers largely eliminate the need for time-varying optical signals and magnetic fields, simplifying control electronics and electro-optics compared with time-domain measurements. Advantages in the noise response of continuous atom interferometers has been demonstrated theoretically \cite{joyet_theoretical_2012}. Gaussian pulse shaping, which occurs automatically in spatial-domain interferometers with Gaussian laser beams, also has advantages in rejection of high-frequency phase noise \cite{fang_improving_2018}.


The simplest scheme for continuous-beam atom interferometry uses a thermal atomic beam from an effusive oven, employing no laser cooling of the atoms \cite{riehle_optical_1991,lenef_rotation_1997, narducci_advances_2021}.
Laser cooling of continuous atomic beams \cite{kwolek_three-dimensional_2020} makes possible increased interrogation time due to the slower mean velocity of cold atom sources, increased static fringe contrast due to the reduction in Doppler width of momentum-changing transitions, and potentially improved performance under dynamics (accelerations and rotations) due to reduction of inhomogeneous broadening of atomic phase and scale factor. Reduction of the atomic velocity width in all three dimensions is important for performance under acceleration and rotation dynamics in three dimensions \cite{lan_influence_2012,black_decoherence_2020, narducci_advances_2021}. 

Laser-cooling in spatial-domain, inertially sensitive atom interferometers has been investigated experimentally. Landmark gyroscope experiments used a fast cesium beam transversely cooled to 2D temperatures above the Doppler limit, while remaining longitudinally hot \cite{gustavson_rotation_2000,durfee_long-term_2006}. A continuous interferometer based on atoms from a low-velocity intense source (LVIS) likewise used transversely cold atoms with a broad longitudinal velocity width and exhibited a resulting small fringe amplitude \cite{xue_continuous_2015}.  Zero-dead-time operation of pulsed, time-domain cold-atom interferometers has been achieved in long-interrogation-time fountain geometries \cite{dutta_continuous_2016}. \hl{The continuous cold-atom fountain clock FOCS-2 is based on a 3D-cooled atom beam, also with a long interrogation time and a highly curved parabolic trajectory \cite{devenoges_improvement_2012}. The long interrogation times achieved in fountain geometries can improve sensitivity, while highly curved atomic trajectories can reduce decoherence due to scattered light. However, such trajectories are dependent on the system maintaining a particular orientation with respect to gravity and require very low platform dynamics.} While cooling of atomic beams for atom interferometry has thus been explored, no experiment has yet demonstrated inertially sensitive atomic interferometry in a continuous atomic beam that has been cooled to \hl{sub-Doppler} temperatures in three dimensions.
\begin{figure*}
	\includegraphics[width=\linewidth]{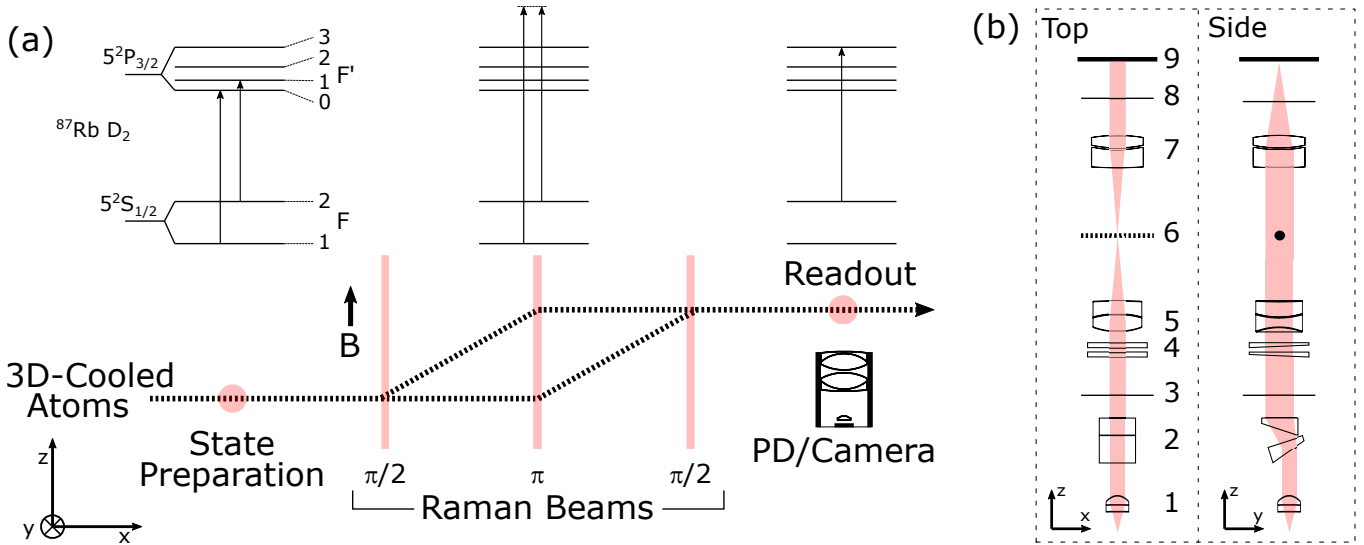} 
	\caption{\label{fig:raman} (a) Inertially sensitive interference is achieved with a 3D-cooled continuous atom beam passing through five regions equally spaced in the $x$ direction: state-preparation; three Raman interactions (pulse areas $\pi/2$, $\pi$, $\pi/2$); and final state-readout. A bias magnetic field B is applied along the Raman beam propagation direction. Not to scale.  (b) A single Raman beam is shown from a top and side view. The beam is launched through a fiber and collimated by an aspheric lens with 11~mm effective focal length (1). The beam travels through a prism-pair (2) which elongates the beam in the vertical direction, and then a $\lambda/4$ waveplate (3) to set the polarization (circular for Doppler-free Raman interactions, linear for Doppler-sensitive Raman interactions). The beam is focused with a 100~mm focal-length cylindrical achromatic doublet (5) in the longitudinal axis of the atom beam (6). The wedge in the cylindrical lens is corrected using a Risley prism pair (4). The beam is collimated in the horizontal axis and focused into a cat's eye in the vertical axis by an air-spaced spherical doublet lens (7) with 100~mm focal length. A second $\lambda/4$ waveplate (8) sets the retroreflected polarization as orthogonal after being reflected from the mirror (9).
	}
\end{figure*}

In this paper, we demonstrate inertially sensitive spatial-domain atom interferometry in a continuously 3D-cooled rubidium-87 beam with sub-Doppler 1D temperatures of 15~$\mathrm{\mu K}$ in all three dimensions. \hl{The experiment does not rely on significant gravitational curvature of the atomic trajectory to reduce decoherence caused by scattered cooling light, making possible operation in arbitrary sensor orientations and under dynamics.} Instead, we make use of a 3D-cooled atom-beam source designed to reduce decoherence induced by scattered light \cite{kwolek_three-dimensional_2020}. \hl{The 3D-cooled atom source is relatively compact, with a length of approximately 10~cm, enabling portable sensor applications.} With three cw laser beams driving Doppler-sensitive Raman transitions, separated by $L=7.2(1)$~$\mathrm{cm}$ with interrogation time $T=6.7$~$\mathrm{ms}$, we achieve interference fringe contrast $C=0.30$. (Here, contrast $0 \leq C \leq 1$ is \hl{$C=\alpha/\bar{V}$}, where \hl{$\alpha$} is the amplitude of the fringe signal and $\bar{V}$ is the signal level at mid-fringe.) We demonstrate inertial sensitivity and validate first-principles scale-factor estimates through comparison with a classical accelerometer-seismometer. Based on noise measurements inferred from data taken outside of the contrast envelope of the inertially sensitive fringe measurement, we estimate interferometer self-noise, including quantum projection noise, leading to a phase measurement noise of 530(20)~$\mathrm{\mu rad/\sqrt{Hz}}$.

We further demonstrate zero-dead-time, spatially resolved phase-shear readout of the continuous atom interferometer by tilting one Raman beam and performing fringe readout on a camera at a rate faster than $1/T$. This method provides high-rate quadrature detection of atomic phase, reducing sensitivity to drifts in atom flux and fringe contrast while increasing dynamic range.

\section{Apparatus}

\begin{figure}
\centering
	\includegraphics[width=\linewidth]{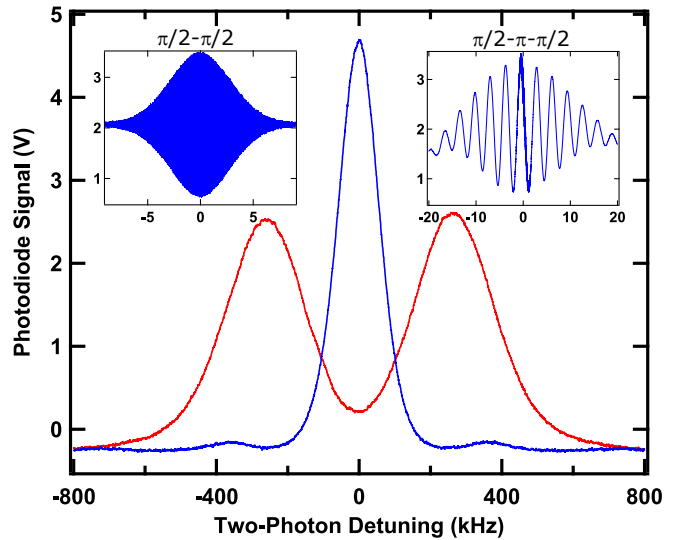}
	\caption{\label{fig:raman_transitions} (Color Online) A single-shot scan, lasting 200~ms and low-pass filtered at 3 kHz, over the Doppler-free (blue, single peak) and Doppler-sensitive (red, two peaks) two-photon Raman transitions is shown for intensities corresponding to a $\pi$-pulse for each respective curve. A fit of the Doppler-free distribution to a Gaussian reveals a $1/e$ radius of 82(1)~kHz, corresponding to a Gaussian Raman beam waist of approximately 40~$\mathrm{\mu}$m assuming an atomic velocity of 10.75~m/s (separately measured through Ramsey fringe spacing). Splitting of the Doppler-sensitive peaks implies an angle of $0.55(1)\deg$ between the atomic beam propagation direction and the Raman beam propagation direction. A fit to the peaks reveals a temperature of 13.4(2)~$\mathrm{\mu}$K after correction for the transit-time broadening indicated by the Doppler-free spectrum. The left and right insets contain two-pulse ($\pi/2-\pi/2$ Ramsey) and three-pulse ($\pi/2-\pi-\pi/2$ echo Ramsey) Doppler-insensitive interferometry sequences respectively. The two-pulse interferometer is performed at beam separation of $14.4\;\textrm{cm}$, and the measured fringe period provides a measurement of mean atomic velocity. The fringes shown in the three-pulse sequence are used to equalize the distances between the first two and last two pulses. The fringes shown here are due to a deliberate distance mismatch of $\Delta L = 550\;\mu\textrm{m}$. \hl{The increased line thickness at the central fringe is due to residual Ramsey fringes at high frequency, which appear due to a small error in pulse area for the three-pulse configuration.}
}
\end{figure}

The configuration of the continuous 3D-cooled atom interferometer is shown in Figure~\ref{fig:raman}. It is based on a previously demonstrated 3D-cooled $^{87}\textrm{Rb}$ beam combining a 2D$^+$ magneto-optical trap and a tilted moving 3D polarization gradient cooling stage \cite{kwolek_three-dimensional_2020}. The key features of the rubidium beam are sub-Doppler temperatures in 3D, dynamically controllable longitudinal velocity (6 - 16 m/s), high flux (greater than $10^9$~atoms/s), and relatively low downstream fluorescence emitted from the cooling stages, resulting in low decoherence in atomic interference measurements made using the cold beam. In our earlier work we describe measurements of the 3D velocity distribution, achieving temperatures of approximately 15~$\mathrm{\mu}K$ along each axis. In the experiments described here, a mean longitudinal atomic velocity of 10.75(20)~m/s is used, as measured from Doppler-insensitive Ramsey fringe period (see Figure~\ref{fig:raman_transitions}) and Raman beam spacing $L$.

To demonstrate inertially sensitive atomic interferometry, we admit the 3D-cooled atom beam into an elongated vacuum cell with five regions, equally spaced in the $x$ direction by 7.2(1)~cm: (1) state preparation, (2) Raman $\pi/2$, (3) Raman $\pi$, (4) Raman $\pi/2$, and (5) detection as seen in Figure~\ref{fig:raman} (a). \hl{Equal spacing between the Raman beams is necessary for interferometer operation, but the spacing of the state preparation and detection regions is somewhat arbitrary.} For clarity we introduce the following orthogonal axes: $x$ is the direction of atomic beam propagation, $y$ is the direction of gravity in this experiment, and $z$ \hl{is perpendicular to $x$ and $y$,} approximately along the direction of Raman beam propagation. The state preparation region (1) utilizes two linearly-polarized laser beams, copropagating along $y$ and retroreflected, tuned to the $F=2\to F'=1$ and $F=1\to F'=0$ D2 transition lines respectively. This prepares \hl{$94\%$ of the} atoms into the $F=1$, $m_F=0$ ground state. The three Raman pulses drive a two-photon transition between the $F=1, m_F=0$ and $F=2, m_F=0$ ground states, spaced by 6.834~GHz. The Raman beams interrogate the atomic beam at a slight angle $\approx 0.5\degree$ from \hl{the $z$ axis, in order to provide a mean Doppler shift that distinguishes between the two directions of photon recoil}. A magnetic field, $B=200$~mG, is applied along the Raman beam propagation direction.

Detection of the atomic state at the output of the interferometer occurs by measuring the fluorescence due to the selective excitation of the $F=2$ ground atomic state by the \hl{circularly polarized} readout laser beam, which propagates along $y$ and is retroreflected. The readout beam is tuned to resonance with the rubidium D2 $F=2 \rightarrow F'=3$ transition. Fluorescence along the $z$ axis is collected via a reentrant window, using a three-lens light collection system.

\begin{figure}[h]
	\centering
	\includegraphics[width=\linewidth]{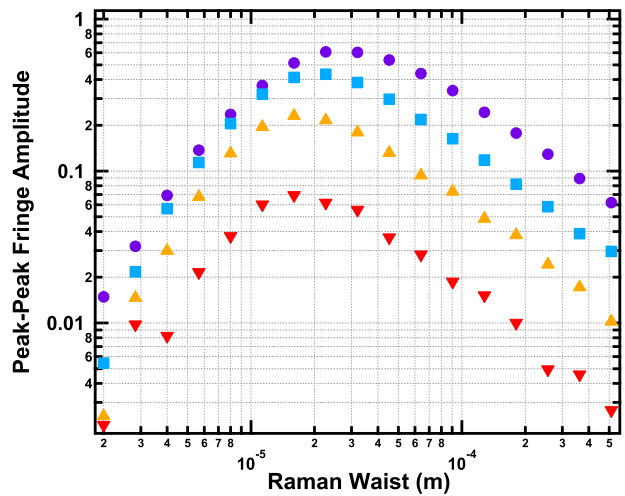}
	\caption{\label{fig:waist} (Color Online) Modeled peak-to-peak fringe amplitude, normalized to total atomic flux, is plotted as a function of Raman beam waist for 3D atom beam temperatures of 3~$\mu$K (purple circles), 15~$\mu$K (blue squares), 75~$\mu$K (yellow triangles), and 500~$\mu$K (red upside-down triangles). Currently, we operate with a beam waist of approximately 40~$\mu$m and 3D temperature 15~$\mu$K, which corresponds to a maximally achievable fringe contrast of 33\%.
	}
\end{figure}

\hl{The required shape of the Raman beam is highly elliptical, due to a compromise between two factors that compete in the optimization of fringe amplitude: the need to interrogate a large fraction of the atomic velocity distribution, and the need for the Raman beam profile to be sufficiently uniform across the atomic position distribution, with relatively flat wavefronts \cite{wicht_phase_2005, Cheng_effects_2014, cervantes_selection_2021}. In a spatial-domain atom interferometer, the Raman pulse's transform-limited bandwidth, and hence the range of Doppler shifts addressed by the Raman pulse, is determined by the size of the Raman beam waist. Thus, the higher the atomic transverse temperature, the narrower the Raman beam waist needed to achieve large fringe amplitude. On the other hand, the smaller beam waist reduces the Rayleigh range compared with the transverse size of the atom beam, and increases the influence of wavefront curvature and Gouy phase. In combination, these effects lead us to choose highly elliptical Raman beams, with long waist along the $y$ direction and narrow waist along the direction nearly parallel to $x$. These considerations are reflected in Figure~\ref{fig:waist}, described below.}

\hl{To evaluate quantitatively the influence of the Raman beam shape on interferometer fringe contrast, we use a numerical two-level model of spatial-domain atom interferometer operation. Numerical solution of the time-dependent Schr\"odinger equation for a two-level atom driven by a field with time-varying Rabi frequency and phase provides the atomic state following Raman beam interactions. The driving field parameters are given by the mode function of an elliptical Gaussian beam, including wavefront curvature and Gouy phase shift. Free evolution time $T$ between Raman beams is determined by Raman beam spacing and atomic velocity. Atomic temperature along the Raman propagation direction is accounted for as a Doppler detuning from two-photon resonance. Atomic positions and velocities are treated classically and the interferometer fringe signal is determined by integration over normal position and velocity distributions. We estimate fringe amplitude at a variety of Raman beam waists along $x$ and atomic temperatures in 3D. The results of this analysis are plotted in Figure~\ref{fig:waist} for our measured transverse atom beam size (rms) $\sigma_\mathrm{atom}=1.5$~mm, and for a Raman beam waist along $y$ of 4.1~mm.}

\hl{At our typical operating temperature of 15~$\mu$K, the optimum Raman beam waist is around 25~$\mu$m, while the beam waist inferred from the Doppler-insensitive Raman transition spectrum in Figure~\ref{fig:raman_transitions} is 40~$\mu$m. The analysis of Figure~\ref{fig:waist} suggests that this limits the fringe contrast to 33\%, slightly higher than the observed best-case fringe contrast of 30(1)\%. We note that this analysis does not incorporate contrast reduction due to spontaneous emission.}

The Raman beams are generated from a single laser blue-detuned by $1.1$~GHz from the $^{87}\textrm{Rb}$ $\mathrm{5S}_{1/2}\to\mathrm{5P}_{3/2}$ $\mathrm{F}=2 \rightarrow \mathrm{F}'=3$ transition. The Raman beams are modulated at approximately $6.834$~GHz using fiber-coupled electro-optic phase modulators (IXBlue NIR-MPX800-LN-10), so that each Raman beam is independently phase modulated. Microwave signals driving the phase modulators are generated by a common 6.734~GHz source (National Instruments QuickSyn FSW-0010) mixed with three direct digital synthesis channels (Wieserlabs FlexDDS-NG) in single-sideband mixers (Polyphase SSB4080A). The reference clocks of the signal sources are provided by a common GPS-disciplined crystal oscillator. The single-sideband mixers allow frequency agility while maintaining relative phase stability between the three Raman beams. \hl{The phase noise between the rf channels was measured to be 400~$\mu$rad, integrated above 1~Hz.} In general, we drive the Raman $\pi$ beam modulator at a small frequency difference, $0 \leq f_R \leq 700$~Hz, from the $\pi/2$ beam modulation. \hl{When $f_R > 0$, the interferometer phase ramps at a rate $\dot{\Phi} = 4 \pi f_R$, resulting in an oscillatory atomic population from which the inertial phase can be determined by lock-in demodulation.} The three phase modulator optical outputs are routed to the experiment using polarization-maintaining patch fiber (PANDA 5/125).

\hl{To create these elliptical Raman beams with relatively flat wavefronts, special care is taken in beam shaping.} Each Raman beam is launched from its fiber and collimated using an aspheric lens (labeled 1 in Figure~\ref{fig:raman} (b)). The beam is sent through an anamorphic prism pair (2) with 3x magnification along the $y$ axis. A cylindrical achromatic doublet (5) with focal length 100~mm focuses the collimated elliptical beam in the $x$ axis, creating an elliptical beam with roughly $100:1$ aspect ratio at the position of the atoms. The resulting Raman beam at the atoms has waists of 4.1~mm along $y$ and $40~\mathrm{\mu m}$ along $x$. Each Raman beam is retroreflected using a combination of a spherical achromatic doublet lens (7) and a planar mirror (9). The lens collimates the Raman beam in $x$ direction and focuses it in the $y$ direction, and the retroreflecting mirror placed at the focus of this lens creates a cat's eye configuration. The total optical power in the beam driving the Raman $\pi$ pulse is approximately 1.5~mW propagating in each direction, while the power in the $\pi/2$ beams is roughly half that. \hl{The exact power of each beam was determined by optimizing the atomic state population produced by each Raman pulse. In the frequency range from 1~Hz to 1~kHz, the temporal fluctuations in Raman beam power are 0.3\% (rms).} A schematic of the Raman beam generation is shown in Figure~\ref{fig:raman} (b). Examples of Doppler-insensitive and Doppler-sensitive Raman transition spectra are plotted in Figure~\ref{fig:raman_transitions}.

To observe contrast in the inertially sensitive interference signal, all three Raman beams and their retroreflections must be mutually parallel within approximately 100~$\mathrm{\mu}$rad. This follows from the observation that a contrast variation by $\pi$ across the atomic beam results in a cancellation of fringe signal when integrating over all atoms, and so the requirement for angle $\Delta \theta_R$ between the Raman beams is $\Delta \theta_{R} < \frac{\pi}{\sigma_{\textrm{atom}} k_{\textrm{eff}}}$. Here, $k_\textrm{eff}$ is the 2-photon wavevector of the Raman beams, and $\sigma_{\textrm{atom}}$ is the rms transverse size of the atom beam. To achieve sufficiently parallel Raman beam alignment, we employ an autocollimator to align the independent Raman retroreflection mirrors to a common optically flat reference, and we align the input light to ensure backward coupling into the input optical fibers. Low-wedge vacuum windows are used, and wedge in the nonfocusing axis of the cylindrical input lens is corrected for each Raman beam using a Risley prism pair. The $x$ components of the Raman wavevectors are made equal by angular adjustment to equalize the  mean Doppler shift of each Doppler-sensitive Raman transition as shown in Figure~\ref{fig:raman_transitions}.

Likewise, the Raman beams must be equally spaced in order for atomic wavepackets to overlap within the coherence length at the end of the interferometer \cite{kellogg_longitudinal_2007}. For a spatial-domain interferometer with mean atomic velocity $v_{\textrm{atom}}$ and length mismatch $\Delta L$, the resulting requirement is $\Delta L < \frac{v_{\textrm{atom}}}{k_{\textrm{eff}} v_\textrm{th}}$, where $v_{\textrm{th}} = \sqrt{\frac{k_B \tau}{m}}$ is the thermal velocity, $k_B$ is the Boltzmann constant, $m$ is atomic mass, and $\tau$ is atomic temperature in the direction along the Raman beams. To implement accurately spaced beams, we perform a Doppler-free spin-echo interferometry sequence with a $\pi/2$, $\pi$, and final $\pi/2$ pulse. The result, when scanned over two-photon detuning, reveals a Gaussian spectral profile with overlaid oscillations due to length mismatch between the first two and the last two pulses as seen in Figure~\ref{fig:raman_transitions}. By modifying the position of the third pulse, the oscillation frequency can be driven down below the detectable limit, which corresponds to a length difference on the order of the spatial extent of the Raman pulse.

\section{Experiment}

\begin{figure*}
\centering 
	\large{\bf \hspace{-.41\linewidth} (a)\hspace{.48\linewidth} (b)} \hfill\\
	\includegraphics[width=.45\linewidth]{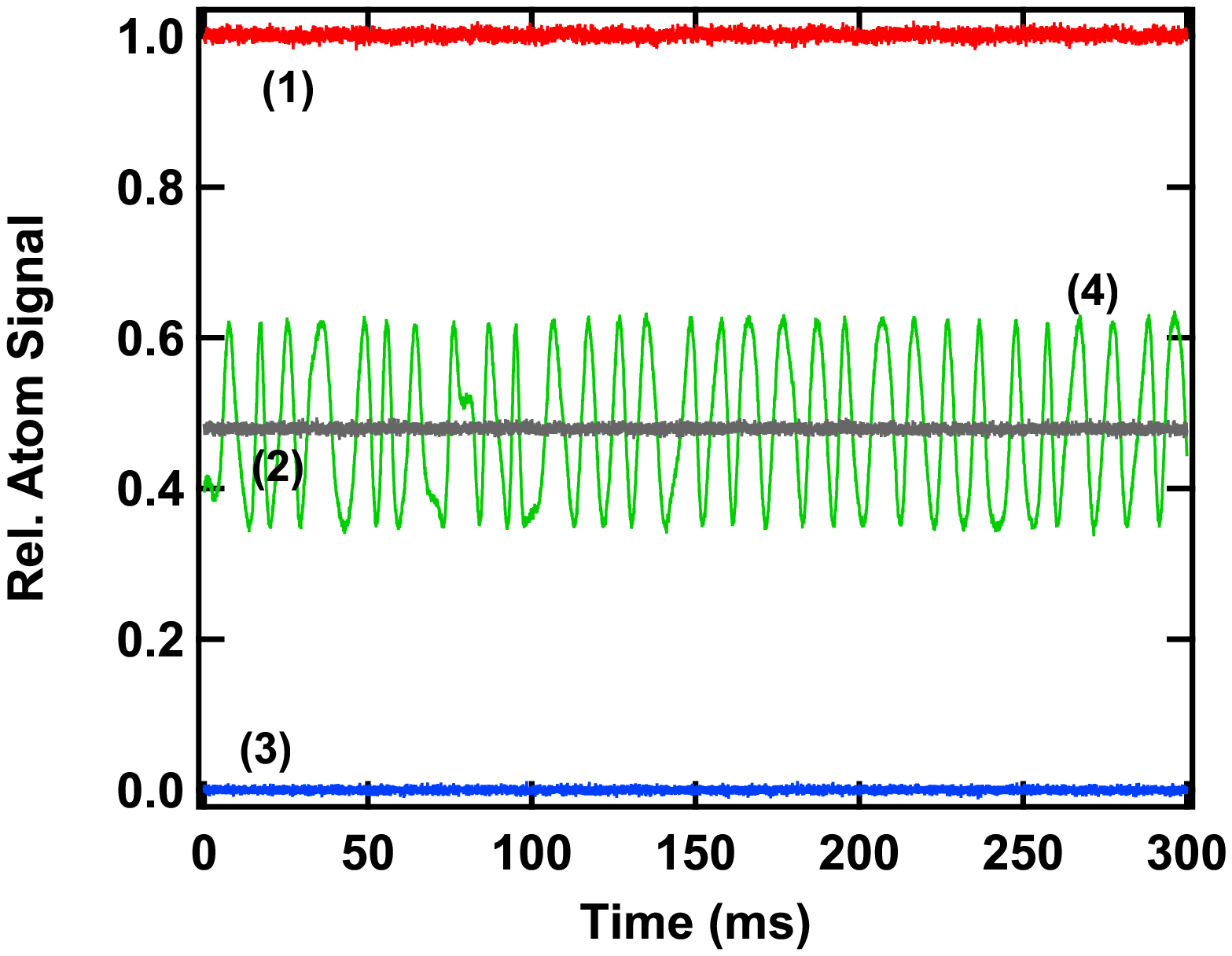}\hspace{.05\linewidth}\includegraphics[width=.45\linewidth]{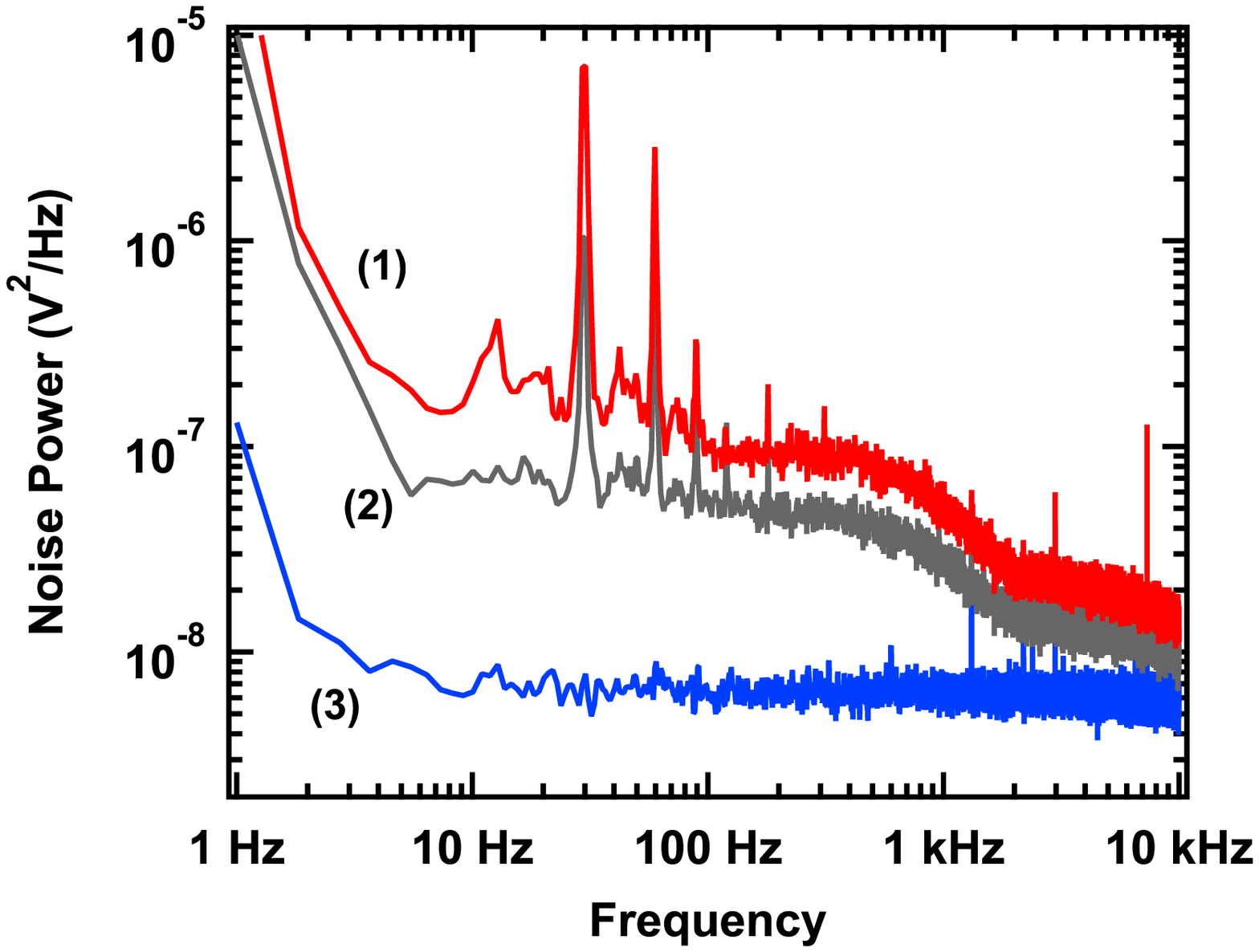}\\
	\large{\bf \hspace{-.41\linewidth} (c)\hspace{.48\linewidth} (d)} \hfill\\
	\includegraphics[width=.48\linewidth]{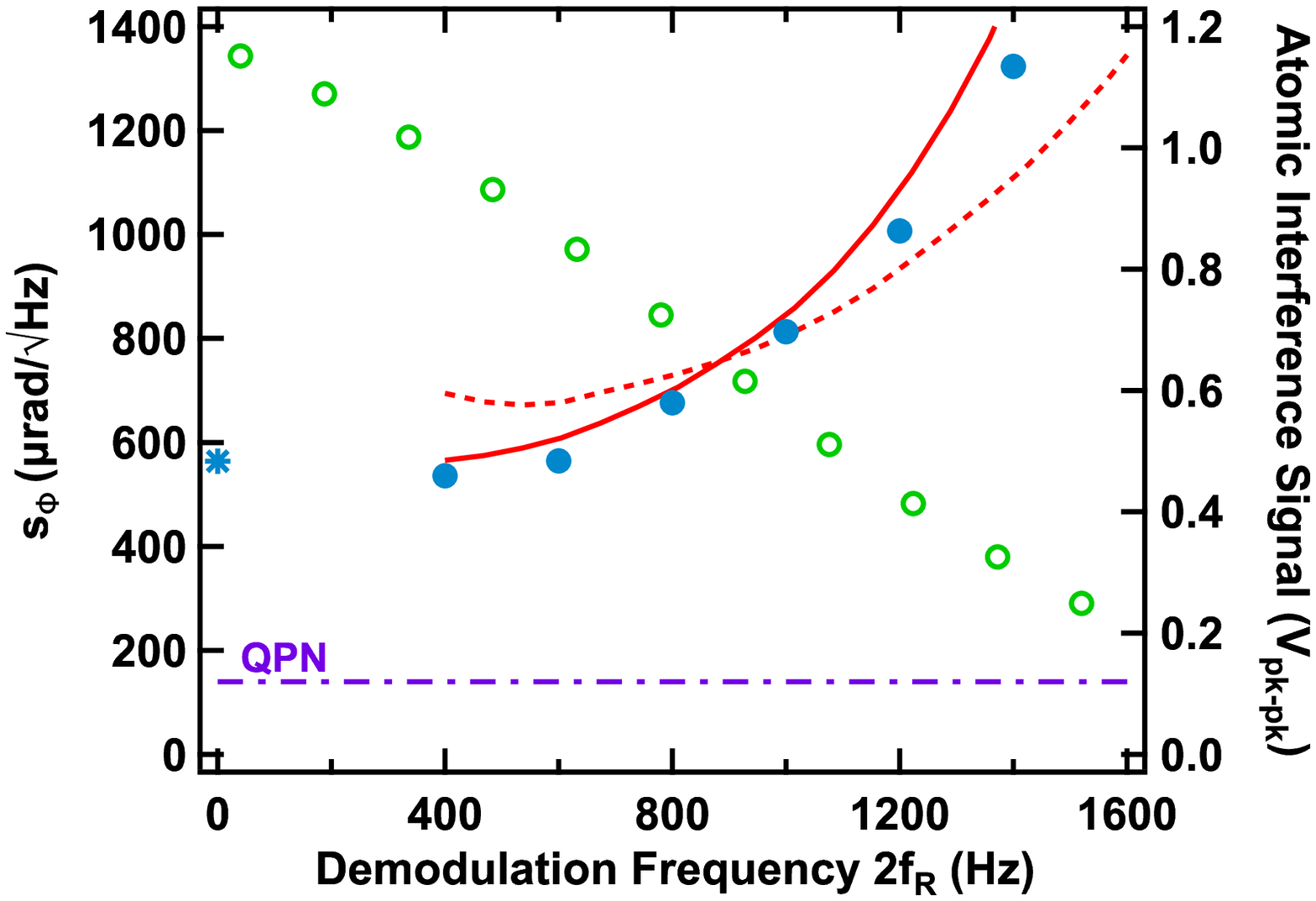}\hspace{0.02\linewidth}\includegraphics[width=.45\linewidth]{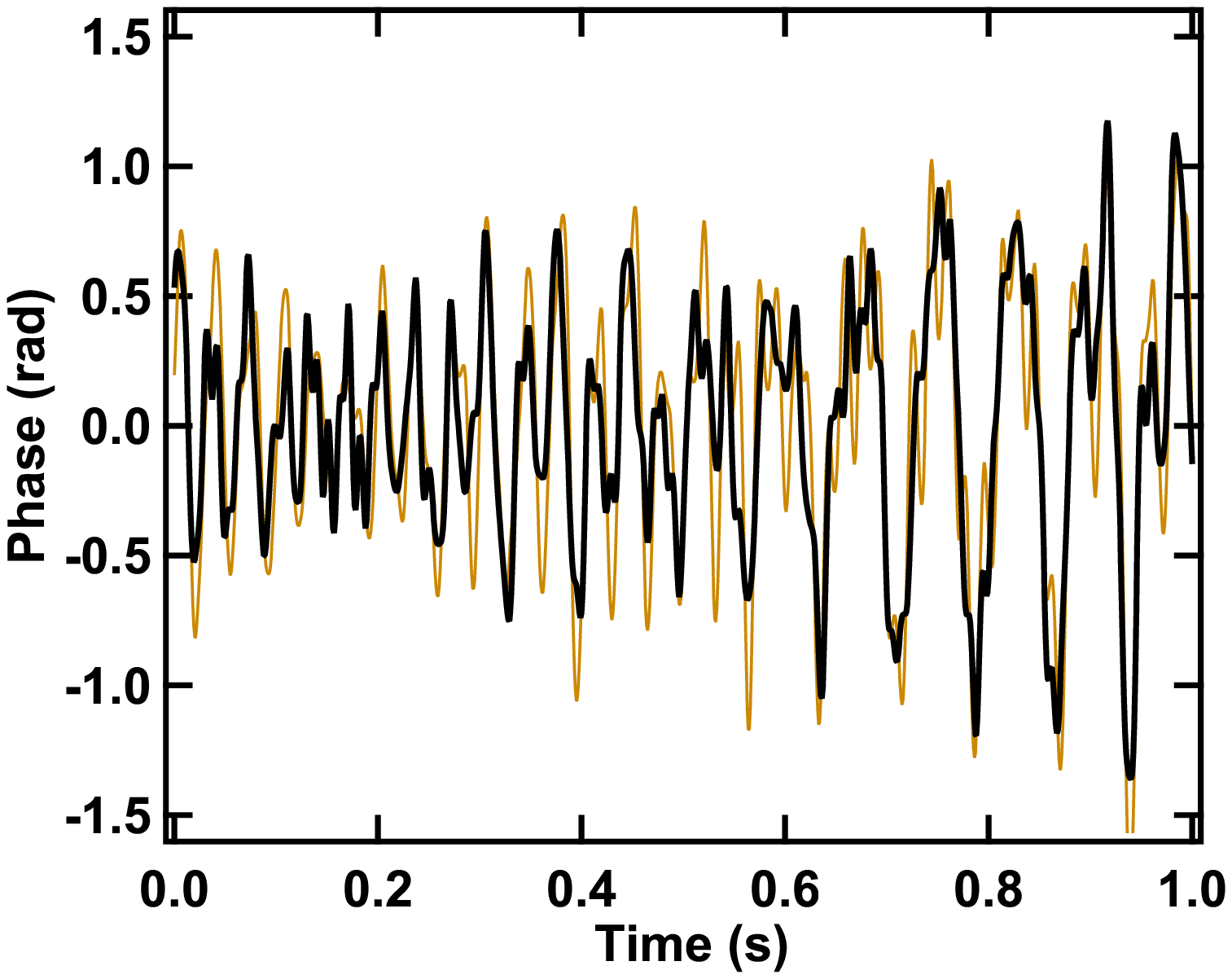}
	\caption{\label{fig:psd} Time series (a) and one-sided power spectral densities (b) of state-selective atomic $F=2$ fluorescence signals are acquired under three different conditions: (1) Red line: total atom flux with Raman beams and optical pumping beam blocked, so that the entire atom beam is in $F=2$. (2) Gray line: Atomic interference signal taken outside of the interference contrast envelope to measure noise by imposing a $f_R=5$~kHz frequency difference between the $\pi$ and $\pi/2$ beams. (3) Blue line: Atom beam turned off by blocking the cooling beams. An additional time trace (4) shows the inertially sensitive atomic interference signal following optical pumping to $F=1$, $m_F=0$ and the Raman beam sequence, with $f_R=50$~Hz. In the inertially sensitive fringe time series, the effects of both the slow Raman modulation phase ramp and platform vibrations are visible. The phase measurement noise $s_\Phi$ of an interferometer (c) is predicted (red lines) and measured experimentally (blue filled circles) with a lock-in amplifier for multiple demodulation frequencies at a 6~dB, 3~ms filter. The concurrent fringe signal $2 \alpha$ is shown (green empty circles), resulting in increasing $s_\Phi$ as a function of $f_R$. The phase noise in (c) is predicted from the noise power spectrum (b) for a detection region size $w_\mathrm{det} = 4.3$~mm (red solid line) and $w_\mathrm{det} = 2.7$~ mm (red dashed line), each with a 3~ms lock-in time constant, as well as a DC-coupled side-of-fringe lock (blue asterisk) with 150-Hz bandwidth. The quantum projection noise limit is shown as a horizontal purple dot-dashed line. (d) Measured atom interferometer phase (thick black line) is compared with a calculated phase based on classical accelerometer measurements (thin brown line) on a vibrationally noisy platform.
	}
\end{figure*}

\subsection{Interference Measurement}
\label{Sec:Interference}
The phase of the interferometer is inferred from the atomic $F=2$ state population in the readout region. State-selective fluorescence is performed with an on-resonance probe on the D2 cycling transition (see Figure~\ref{fig:raman} (a)). An avalanche photodiode converts detected power into a voltage, which is amplified and low-pass filtered at 10~kHz. A typical peak-to-peak fluorescence signal of the inertially-sensitive interferometer is approximately \hl{$2\alpha=1.3$~V}. The highest fringe contrast observed in day-to-day operation is $C=0.30(1)$, with a mean signal level $\bar{V}$ approximately equal to 50\% of the full atomic flux signal size as shown in Figure~\ref{fig:psd}(a).

The fluorescence signal can be converted into an inertially sensitive phase either at DC (identical Raman modulation frequencies, $f_R=0$), or through the imposition of a Raman modulation frequency difference $200~\mathrm{Hz}~\leq~f_R~\leq~700~\mathrm{Hz}$ and phase detection through lock-in demodulation at $2 f_R$. Practical operation at DC would employ feedback control of phase modulator drive signals to stabilize interferometer phase at the point of maximum signal slope. Here we evaluate noise for DC operation, but we primarily operate at $f_R > 0$ for the experiments described in this section.

For lock-in detection, a lock-in amplifier is operated with reference frequency $2 f_R$ to measure the two quadrature components (sine and cosine) of the atom signal, allowing simultaneous estimation of fringe amplitude and phase over a range of $\pm \pi$, \hl{similar to the methods of Gustavson et al. \cite{gustavson_rotation_2000}}. Operation at $f_R>0$ reduces the influence of $1/f$ and other low-frequency noise as shown in Figure~\ref{fig:psd}(b). Due to finite detection beam size and finite atom temperature in the spatial-domain atom interferometer, fringe contrast is a decreasing function of $f_R$, as plotted in Figure~\ref{fig:psd}(c). For a Gaussian-weighted detection region of $1/e^2$ radius $w_\mathrm{det}$ along the direction of atomic propagation, measured fringe contrast is proportional to $\exp{(-4 \pi^2 w_\mathrm{det}^2 f_R^2 /(2 v_\mathrm{atom}^2))}$ in the limit of zero atomic temperature. The effect of finite atomic temperature may be determined numerically. \hl{While the typical experiment conducted here used a detection beam with radius $w_\mathrm{det}=4.3$~mm, the size which optimizes interferometer performance changes depending on the demodulation frequency used, as shown in Figure~\ref{fig:psd} (c).}

Phase variation in the inertially sensitive mode of operation is almost entirely due to true vibrational acceleration and rotation of the measurement platform, making it challenging to measure the interferometer's intrinsic noise level. We therefore adapt the method of Gustavson et al. \cite{gustavson_rotation_2000}, estimating noise on the measurement of atomic phase by a rotation- and acceleration-independent measurement of the interferometer signal obtained following a sequence of three $\pi/2$, $\pi$, and $\pi/2$ Doppler-sensitive Raman pulses with large $f_R = 5$~kHz. In this noise estimation technique, the modulated phase varies so quickly that our phase detection technique observes no fringe contrast and only detects noise.

Typical one-sided power spectral densities (PSDs) of measured noise are shown in Figure~\ref{fig:psd} (b) under three different conditions: full atomic flux without application of optical pumping or Raman beams, a zero-contrast interferometer as described in the preceding paragraph, and zero atomic flux representing noise in the detection system. The noise spectra reveal that the dominant noise appears on the full atomic flux detection, with peaks near 30~Hz and subsequent harmonics due to vibrational noise in the 2D MOT and 3D optical molasses optical alignment. Integration of the noise PSD in the case of $f_R = 5$~kHz indicates a noise level for the inertially sensitive measurement. To predict real-world performance, we model different methods of phase-sensitive readout and estimate the resulting phase noise. For DC-phase detection, the amplitude spectral density (ASD, or root-PSD) $s_\Phi$ of phase measurement noise is directly estimated from ratio of the ASD of zero-contrast atomic fluorescence $s_{NC}$ and the inertially sensitive fringe amplitude $A$ (measured for $f_R=0$) by $s_\Phi = s_{NC} / A$. For near-DC phase measurement, we thus estimate the mean noise density in the measurement band from 1~Hz to 150~Hz to be $560(20)~\mathrm{\mu rad / \sqrt{Hz}}$.

For noise analysis of lock-in detection of interferometer phase, we mimic the transfer function of a lock-in amplifier with a 6 dB/octave filter with 3 ms time constant, and integrate over the PSD to determine the predicted phase measurement noise for $f_R \geq 200$~Hz. Lower values of $f_R$ are not evaluated here in order to allow the filter time constant to remain high enough to measure accelerations and rotations at frequencies up to $1/T = 149$~Hz while still filtering out the carrier at $2 f_R$. We validate this measurement using lock-in detection as plotted in Figure \ref{fig:psd} (c). For the present size of the detection region along $x$, the lowest inferred phase measurement noise occurs at $f_R = 200$~Hz, with mean noise density equaling $530(20)~\mathrm{\mu rad / \sqrt{Hz}}$. In Section~\ref{Sec:Discussion} below, we estimate the sensitivity of dual-beam accelerometer and gyroscope sensors with this level of intrinsic phase measurement noise.

The  effective size of the state detection region along the $x$ direction is $w_\mathrm{det}=4.3$~mm including the effects of detection beam, demagnification, and photodetector size, so that fringe contrast as a function of $f_R$ is reduced by a factor of $e$ at $f_R = 550$~Hz. We additionally measure noise spectra at a smaller detection beam size  of $w_\mathrm{det} \approx 2.7\;\textrm{mm}$, corresponding to a smaller signal but also enabling higher values of $f_R$ with $1/e$ contrast dropoff of 850~Hz. The estimated phase measurement noise for the smaller detection beam is plotted as a dashed line in Figure~\ref{fig:psd} (c). Reduction in the detection beam size improves signal-to-noise ratio at higher modulation frequencies.

\subsection{Classical Accelerometer Correlation}
The phase of the inertially sensitive interferometer is predominantly due to platform accelerations. We perform a simultaneous lock-in measurement of atom interferometer phase and acceleration measurement on two classical accelerometers (Wilcoxon 731A) bolted to the atom interferometer platform, spatially separated by approximately 1 meter along $x$, with sense axes parallel to the Raman beam direction. This allows us to demonstrate the response of the atom interferometer to motion, and to show that the classical accelerometer noise level is adequate to perform feed-forward phase compensation of the atom interferometer to disambiguate $2 \pi$ phase wraps of the atom interferometer under platform dynamics.

Because of the differing measurement bandwidths of the atom interferometer and classical accelerometers, for the sake of comparison we use the classical acceleration signals to infer a calculated atom interferometer phase, which is displayed along with the measured atom interferometer phase in Figure~\ref{fig:psd}(d). To infer phase from the classical interferometer measurements, we integrate the acceleration signals of the two accelerometers to obtain accelerometer positions, and assume rigid-body motion of the platform to infer the positions of the Raman retroreflection mirrors versus time. \hl{For small-amplitude accelerations and rotations, the Raman mirror positions are estimated by linear interpolation of the accelerometer positions.} Based on these calculated mirror positions, atom interferometer phase is predicted. The manufacturer specified scale factor for the accelerometers is used, along with the independently known values for the atom interferometer's atomic velocity, Raman beam separation, and Raman wavevector. While no arbitrary scale factor is introduced, a small overall time delay between the \hl{calculated and measured interferometer phase} is adjusted slightly to account for signal propagation delays.

Hybridization of an atom interferometer with classical inertial sensors can resolve the inherent $2 \pi$ phase ambiguity of atom interferometers, thus increasing dynamic range enormously \cite{lautier_hybridizing_2014,narducci_advances_2021}. This can be accomplished by using a classical inertial sensor output to drive a feed-forward to Raman beam frequency and phase. In order for this feed-forward to resolve $2 \pi$ phase ambiguity in the atom interferometer signal, it is necessary for the classical accelerometer's phase prediction to be accurate within $\pi/2$. The vibrationally noisy platform has rms acceleration of $150~\mathrm{\mu g}$ along the $z$ axis, resulting in rms atom interferometer phase of 600~mrad. (These rms values are not related by the standard phase relation $\Phi = k_\mathrm{eff} a T^2$ because of the differing frequency response of the classical accelerometer and the atom interferometer.) Over a 100~s simultaneous measurement, the rms difference in measured atom interferometer phase and inferred phase based on the classical accelerometers is 277 mrad, while the absolute maximum observed difference is 1.2 rad. \hl{Contributions to this difference include vibrations of the individual Raman mirrors, error in the accelerometer measurements, and measurement axis misalignment.} At least under these relatively static conditions, the classical accelerometer appears to be capable of resolving atom interferometer phase ambiguity. Under greater dynamics, scale factor nonlinearity and cross-axis coupling of the classical accelerometers must be considered.

\subsection{Phase-shear Measurement}

The cold-atom beam interferometer architecture demonstrated here is amenable to the use of a number of advanced or novel atom interferometry techniques. These include rapid atomic velocity switching for high-dynamic-range composite-fringe operation \cite{avinadav_composite-fringe_2020}, actuation of Raman beam alignment to compensate for platform rotation \cite{lan_influence_2012}, rapid k-reversal for cancellation of systematic error with high bandwidth, and continuous spatially resolved point-source interferometry for multidimensional rotation sensitivity \cite{dickerson_multiaxis_2013,avinadav_rotation_2020}. Here we provide an example of one such technique, phase shear interferometry, implemented in the continuous cold-atom beam.

 Phase shear readout has been demonstrated previously in pulsed cold-atom interferometers \cite{sugarbaker_enhanced_2013}, but not (to our knowledge) in continuous-beam interferometers. It provides  spatially resolved fringe measurements that have the advantage of simultaneous measurement of fringe contrast, phase, and background in a single image acquisition. Therefore, fluctuations in atom number or background light levels do not contribute error to the phase estimation. As in lock-in measurement, dynamic range in phase shear readout is increased by a factor of 2 compared with unmodulated, open-loop, side-of-fringe phase measurement. Unlike lock-in measurement, phase shear measurement does not require a low-pass filter to eliminate carrier frequencies since the effective ``carrier" is in the spatial domain rather than the time domain. However, the nonzero camera exposure time does provide an effective upper bound on frequency response.

In the atom interferometer measurements described heretofore, the wavevectors of the three Raman beams are very nearly parallel. The imposition of an angle $\delta \theta$ about the $x$ axis to the final Raman beam causes a spatially dependent atom interferometer phase $\delta \Phi = \delta \theta k_{\mathrm{eff}} y$ to the interferometer output. To demonstrate phase shear interferometry, we rotate the final Raman beam by an angle anywhere from 100 to 800~$\mathrm{\mu rad}$, resulting in different fringe periods appearing in a spatially resolved image of the interference at the interferometer output. Due to the cat's eye retroreflection scheme, the beam remains retroreflected regardless of this small change in input angle, and overlap of the counterpropagating fields is minimally perturbed. \hl{We record phase shear readout images using a CMOS camera (FLIR BFS-U3-51S5M-C) that acquires images in global shutter mode and transmits the images to a computer for analysis.} Typical spatial fringe contrast for our phase shear measurements is 20\%, somewhat lower than observed in time-series of fringes. This limitation could be due to limited depth-of-field of the imaging system, an angular mismatch between the camera's imaging axis and the fringe planes, or mirror vibrations with shorter periods than the imaging exposure time.

\begin{figure}
\centering
	\includegraphics[width=\linewidth]{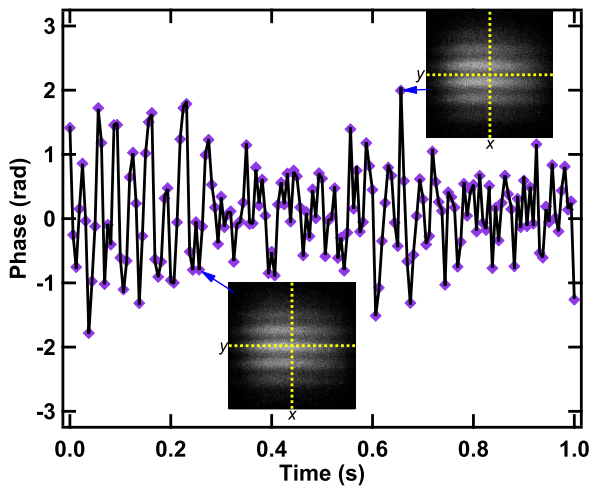}
	\caption{\label{fig:phaseshear}{High-bandwidth phase shear readout is performed by tilting the final Raman beam 400~$\mathrm{\mu rad}$ about the $x$ axis, and imaging the resulting atom beam along the $z$ direction at a rate of 160~Hz, with 5~ms camera exposure time. A fit to the fringes determines the phase and contrast of the interferometer for each acquired image. The resulting inertially sensitive phase is plotted, exhibiting the dominant platform accelerations between 10 and 40 Hz. Black lines show fit result from even image pixel masks, while purple diamonds are fit results from odd image pixel masks (see text). Insets show two example phase shear readout images, with the $x$ axis horizontal and $y$ axis vertical. The imaged areas are 6.5~mm by 6.5~mm. Central crosshairs provide a reference to the eye.
	}}
\end{figure}

Figure \ref{fig:phaseshear} shows an example of inertially sensitive phase shear measurements \hl{acquired} at a rate of 160~Hz, with a final Raman beam angular deviation of $\delta \theta = 400$~$\mathrm{\mu rad}$ about the $x$ axis. A single-image exposure time of 5~ms is used. For each image, the fringe contrast and phase are inferred through \hl{post-processed} fits to sinusoidally modulated 2D Gaussian functions. Insets to the figure show two examples of phase shear images with phases that differ by 2.8~rad.

To evaluate noise in phase estimation by phase shear readout, we adopt the technique of Dickerson et al. \cite{dickerson_multiaxis_2013}: we impose two different checkerboard-pattern masks on the fringe image, offset by one pixel from one another (the ``even" mask and the ``odd" mask) and perform independent phase fits for the even- and odd- masked images, finding the difference between phases for each pair. To reduce the effect of periodic noise, we analyze images filtered by NxN pixel binning, where N=1,2,4. In a set of 1,500 consecutive phase images acquired at 160 Hz, we find that the root-mean-square difference between odd- and even-masked phase fits is 9 mrad for all values of $N$. Under the assumption of uncorrelated, white phase measurement noise, this would imply a phase measurement noise density of $s_{PS}=500~\mathrm{\mu rad/\sqrt{Hz}}$ for the full, unmasked images. This noise level is similar to the noise estimates determined from photodiode-based measurements.

\section{Discussion}
\label{Sec:Discussion}
The estimates of phase measurement noise described above occur in a measurement configuration that is sensitive to noise in atomic flux and fringe contrast, Raman pulse area, detection beam intensity and frequency, as well as photon shot noise and atomic quantum projection noise. However, these noise measurements are insensitive to noise in optical and microwave phase, such as that caused by mirror vibrations or microwave oscillator noise, and therefore represent an underestimate of total interferometer self-noise. Because the platform is vibrationally noisy, inertially sensitive phase measurements [Figures \ref{fig:psd}(a), \ref{fig:psd}(d), and \ref{fig:phaseshear}] predominantly reveal a true acceleration spectrum rather than sensor self-noise. A more complete picture of the sensor's inertial sensitivity can be obtained by measuring in a vibrationally quiet environment or by intercomparison with other highly accurate inertial sensors.

The lowest phase measurement noise inferred by lock-in measurements, approximately $s_\Phi=530~\mathrm{\mu rad/\sqrt{Hz}}$ at $f_R = 200~\mathrm{Hz}$, is somewhat higher than the quantum projection noise (QPN) limit for an atomic flux of $\dot{N} = 10^9$~atoms/s and fringe contrast of $C=0.3$: $s_\mathrm{QPN}=\sqrt{2/\dot{N}}/C \approx 150~\mathrm{\mu rad / \sqrt{Hz}}$ where $s_\mathrm{QPN}$ is again defined as the square root of the one-sided QPN PSD. The similarity of the observed zero-contrast noise spectrum to the noise spectrum of the full atomic flux measurement, Figure~\ref{fig:psd}(b), suggests that technical noise in atomic flux and fluorescence detection are likely the dominant contributions to noise. Improvement in the mounting of cooling beams and the addition of intensity servos will reduce these noise contributions. Additionally, the employment of simultaneous dual-state detection \cite{biedermann_low-noise_2009} will reduce the contributions from atomic flux and detection beam instabilities and is likely to significantly improve phase measurement noise.

If the present interferometer is implemented in a dual-beam accelerometer/gyroscope configuration with the parameters used herein, and assuming identical, uncorrelated interferometer phase measurement noise at the lowest level measured in these experiments, the projected sensitivity to rotations (angle random walk) is given by $ARW = \frac{s_\Phi/\sqrt{2}}{2 \sqrt{2} k_\mathrm{eff} v_\mathrm{atom} T^2} = 17~\mathrm{nrad / \sqrt{s}}$ while the projected sensitivity to accelerations (velocity random walk) is $VRW =  \frac{s_\Phi/\sqrt{2}}{\sqrt{2} k_\mathrm{eff} T^2} = 37~\mathrm{ng / \sqrt{Hz}}$. The factor of $1/\sqrt{2}$ in the numerator arises from the conversion from one-sided root-PSD to Allan deviation according to the standard convention \cite{noauthor_ieee_nodate}, while a factor of $\sqrt{2}$ in the denominator arises from the dual-interferometer measurement assuming uncorrelated noise. If the technical improvements described above make possible a QPN-limited measurement, then the projected sensitivities become $ARW_\mathrm{QPN} = 5~\mathrm{nrad / \sqrt{s}}$ and $VRW_\mathrm{QPN} = 10~\mathrm{ng / \sqrt{Hz}}$.

\hl{Operation of the 3D-cooled atom beam interferometer in a strapdown configuration on a dynamic platform will require further improvements in temperature or reductions in interrogation time compared with the current demonstration. However, the necessary changes in parameters are not extreme: as noted in \cite{black_decoherence_2020}, an interferometer with mean atomic velocity 30~m/s, Raman beam separation 7.5~cm, and 3D atomic temperature $4~\mathrm{\mu K}$ can operate at a rotation rate over 0.25 rad/s and acceleration over 20~$\mathrm{m/s^2}$ along any axis. The phase readout methods demonstrated herein are amenable to operation in dynamic environments: lock-in detection with $f_R >= (2 T)^{-1}$ allows for changes in inertial signal at frequencies up to the first zero in the interferometer's transfer function \cite{cheinet_measurement_2008}, while phase shear readout at high rate in real time is possible through the use of optical intensity masks combined with single-pixel imaging, or else linear overlap integral computation in a fast processor rather than nonlinear curve fitting. Finally, a large number of techniques for hybrid \cite{lautier_hybridizing_2014, cheiney_navigation-compatible_2018}, closed-loop \cite{joyet_theoretical_2012}, or composite-fringe \cite{avinadav_composite-fringe_2020} interferometer operation have been developed to extend dynamic range and are applicable to continuous cold-beam interferometry.}

In this work, we have not addressed the issue of bias stability of the interferometer phase. Indeed, the current design is not expected to exhibit highly stable phase because the Raman beams pass through independent vacuum windows and lenses, and retroreflect from independent mirrors. Achieving stable interferometer phase will require the use of a retroreflection scheme with stable optical path lengths or optical phase locking of the Raman beam paths. \hl{There are reasons to anticipate that the 3D-cooled atomic beam interferometer architecture may be compatible with lower bias and scale factor instability compared with hot-beam interferometers: the interferometer baseline $L=v_\mathrm{atom} T$ is shorter for a given value of $T$, making relative Raman optical path stability easier to maintain; and the stability of $v_\mathrm{atom}$ imparted by the 3D moving polarization gradient cooing stage may be superior to velocity stability in effusive oven sources. These potential advantages require additional study to verify.}

\section{Conclusions}

We have demonstrated the operation of an inertially sensitive, spatial-domain atom interferometer based on a continuously 3D-sub-Doppler-cooled atomic beam. This work demonstrates that a fringe contrast of 30\% can be maintained despite continuous 3D cooling in a geometry without significant trajectory curvature. These demonstrations pave the way for continuous, high-sensitivity acceleration and rotation rate measurements using sensors that can operate in any orientation and under dynamics.

\hl{The benefits of the present approach include high fringe contrast and measurement bandwidth, zero measurement dead time, and an advantageous response to phase noise.} We have demonstrated low inferred phase measurement noise based on out-of-envelope interferometer measurements. Noise levels remain higher than the estimated quantum projection noise limit based on fringe contrast and atomic flux, suggesting that improvements to sensitivity are achievable through improved laser intensity and frequency stability, improved laser pointing stability, and the employment of improved measurement techniques like spatially separated normalized state detection \cite{biedermann_low-noise_2009}. Finally, we have performed the first demonstration of spatially resolved phase-shear readout for a continuous, spatial-domain interferometer.

The single interferometer demonstrated here, while sensitive to motion, does not distinguish between phase due to acceleration and rotation. A dual interferometer employing a second, counterpropagating cold atom beam interrogated by the same set of Raman beams can generate independent acceleration and rotation signals. Further reductions in atomic temperature combined with a modest increase in atomic velocity can ensure that operation on dynamically rotating and accelerating platforms is achievable \cite{black_decoherence_2020}. \hl{Continued advances in the development of cold-atom beams, including continuous beams of degenerate atoms \cite{chen_continuous_2021}, may open up the path to new architectures for atom interferometers benefiting from narrow velocity distributions and continuous operation.}

\section{Acknowledgments}
This work is supported in part by the Office of Naval Research. J.M.K. acknowledges support from a U.S. Naval Research Laboratory Karles Fellowship. Also, we would like to acknowledge the support of Russell Bradley, NRL, with the phase-noise measurements.


\end{document}